\documentclass{aa}

\usepackage{times}
\usepackage{graphics}

\newcommand{\kmprs}  {\mbox{\rm km\,s$^{-1}$}}

\newcommand{\feh} {\mbox{\rm [Fe/H]}}

\newcommand{\teff}  {\mbox{$T_{\rm eff}$}}

\newcommand{\logg}  {\mbox{{\rm log}\,$g$}}

\newcommand{\Lione} {\ion{Li}{i}}
\newcommand{\Lisix} {\element[][6]{Li}}
\newcommand{\Liseven} {\element[][7]{Li}}
\newcommand{\Kxxx} {\element[][39]{K}}
\newcommand{\Kyyy} {\element[][41]{K}}
\newcommand{\sixseven} {\element[][6]{Li}/\element[][7]{Li}}

\newcommand{\Kone} {\ion{K}{i}}
\newcommand{\Caone} {\ion{Ca}{i}}
\newcommand{\Feone} {\ion{Fe}{i}}

\newcommand{\vsini}   {\mbox{$v_{\rm rot}{\rm sin}i$}}

\newcommand{\mvlgt} {\mbox{$M_{\rm V}$-$\log T_{\rm eff}$}}

\begin{document}

%MA \thesaurus{07(08.01.1, 08.01.3, 08.06.3, 08.09.3, 09.03.2)}
\thesaurus{07(08.01.1, 08.01.3, 08.06.3, 08.09.2 G271-162, 09.03.2)}

\title{The lithium isotope ratio in the metal-poor halo star
G271-162 from VLT/UVES observations
\thanks{Based on public data released from the UVES commissioning at the
VLT/Kueyen telescope, ESO, Paranal, Chile}}

\author{P.E. Nissen \inst{1} 
\and M. Asplund \inst{2} 
\and V. Hill \inst{3} \and S. D'Odorico \inst{3}}

\offprints{P.E. Nissen}

\institute{
Institute of Physics and Astronomy, University of Aarhus, DK--8000
Aarhus C, Denmark
(pen@ifa.au.dk)
\and 
Uppsala Astronomical Observatory, Box 515, SE--751 20, Sweden
(martin@astro.uu.se)
\and 
European Southern Observatory, Karl-Schwarzschild Str. 2,
D--85748 Garching b. M\"{u}nchen, Germany
(vhill@eso.org, sdodoric@eso.org)
}

\date{Received March 15 2000/ Accepted April 12 2000}

\maketitle

\begin{abstract}
A high resolution ($\lambda/\Delta\lambda \simeq 110\,000$), 
very high $S/N$ ($\ga 600$)
spectrum of the metal-poor turnoff star \object{G\,271-162}
has been obtained in connection with the commissioning of 
UVES at VLT/Kueyen. Using both 1D hydrostatic and 3D hydrodynamical
model atmospheres, the lithium isotope ratio has been estimated
from the \Lione\,670.8\,nm line by means of spectral synthesis.
The necessary stellar line broadening (1D: macroturbulence + rotation,
3D: rotation) has been determined from unblended \Kone, \Caone\ and
\Feone\ lines. The 3D line profiles agree very well with the
observed profiles, including the characteristic line asymmetries. 
Both the 1D and 3D analyses reveal a possible detection
of \Lisix\ in \object{G\,271-162}, $\sixseven = 0.02\pm0.01$ ($1\sigma$).
It is  discussed if
the smaller amount of \Lisix\ in \object{G\,271-162} than in the similar
halo star \object{HD\,84937} 
could be due to differences in stellar mass and/or metallicity or 
whether it may reflect an intrinsic scatter of \sixseven\ in the ISM
at a given metallicity.

%MA \keywords{Stars: abundances -- Stars: atmospheres -- Stars: fundamental
% parameters -- Stars: interiors -- cosmic rays}

\end{abstract}

\section{Introduction}

Due to the special status of \Lisix\ 
for astrophysics and cosmology, 
much work has been devoted to the search for this isotope in metal-poor stars 
ever since the first detection
in \object{HD\,84937} by Smith et al. (\cite{smith93}). 
The reason for this interest is threefold:
{\em i)} The presence of \Lisix\ in the envelope of 
halo stars severely limits the possible
depletion of \Liseven, and thus allows a more accurate determination
of the primordial \Liseven\ abundance 
(Ryan et al. \cite{ryan99}; Asplund \& Carlsson \cite{asplund00});
{\em ii)} \Lisix\ abundances provide an additional test of theories
for the production
of the light elements (Li, Be and B) by cosmic ray processes;
{\em iii)} Since \Lisix\ is an even more fragile nuclei than \Liseven\,
it is a sensitive probe of possible mixing events 
during the stellar life.

To date \Lisix\ is claimed to have been detected in two metal-poor halo
stars, \object{HD\,84937} and \object{BD\,+26 3578}, and two metal-poor
disk stars, \object{HD\,68284} and \object{HD\,130551}
(Smith et al. \cite{smith98};
Hobbs \& Thorburn \cite{hobbs97}; Cayrel et al. \cite{cayrel99};
Nissen et al. \cite{nissen99}). Clearly, an observational test of 
models for the formation and evolution of \Lisix\ and for the depletion
of \Lisix\ in stellar envelopes requires a much larger set of \Lisix\ data
spanning a large metallicity range. With the advent of high-resolution
spectrographs on 8m-class telescopes this is now becoming feasible.

The determinations of \Lisix\ abundances are based on the increased
width and asymmetry of
the \Lione\ 670.8\,nm doublet introduced by the isotope shift 
(0.16\,\AA\ = 7.1\,\kmprs) of \Lisix. Since the line is not resolved,
the derived \Lisix\ abundance 
depends on the adopted stellar line broadening as estimated
from other spectral lines. Until now all such analyses have
relied on 1D hydrostatic model atmospheres which cannot 
predict the inherent line asymmetries introduced by the convective
motions in the atmosphere and thus is a potential source of
uncertainty. An attractive alternative is provided by
the new generation of 3D hydrodynamical model atmospheres 
(e.g. Stein \& Nordlund \cite{stein98}; Asplund et al. \cite{asplund99},
\cite{asp00}) which
self-consistently compute the time-dependent convective velocity
fields and thus are able to predict the convective line asymmetries.

In the present {\em Letter} we analyze a high resolution,
very high $S/N$ spectrum
of the metal-poor halo star \object{G271-162} obtained during the commissioning
of UVES on VLT/Kueyen using both 1D and 3D model
atmospheres. Thereby we also investigate
the ability of UVES to provide high quality spectra as needed in many
astrophysical studies.
%Thereby we also investigate the aptitude of UVES
%in connection with stellar astrophysics studies that depend on the availability
%of high resolution and very high $S/N$ spectra.

\section{Observations and reductions of UVES spectra}
The metal-poor star \object{G\,271-162} (\object{BD\,$-$10 388})
was chosen as one of the targets
for the first commissioning of UVES,
the UV-Visual Echelle Spectrograph 
(D'Odorico \& Kaper, \cite{dodorico00}) on the ESO VLT UT2 (Kueyen)
telescope. The aim was to test if UVES can supply high quality spectra
suitable for a determination of the lithium isotope ratio in stars
-- one of the most demanding problems in observational stellar astrophysics.
Being situated at the turnoff point for halo stars in the HR diagram,
\object{G\,271-162} is very similar to \object{HD\,84937};
hence, one would expect the Li isotope ratio
to be about the same, i.e. $\sixseven \simeq 0.06$.
Furthermore, G\,271-162 
represents a group of about 50 halo turnoff stars
with $10 \la V \la 11.5$
that are too faint to be searched for \Lisix\ using high resolution 
spectrographs at 4m class telescopes. With UVES it should be possible
to reach these stars and hence to increase the sample of stars
with \Lisix\ determinations by an order of magnitude.

Four spectra of \object{G\,271-162} were obtained on October 11 - 13, 1999,
with exposure times ranging from 30 to 60\,min.  The width of the
entrance slit was set at 0.3 arcsec to get a resolution close to 120\,000.
An image slicer is foreseen for this high resolution mode of UVES,
but was not available during the first commissioning run. Instead,
the star was continuously moved along the slit within $\pm 5$ arcsec
corresponding to approximately 50 pixels on the CCDs. This widening of
the echelle orders is essential in obtaining very high S/N.
The lack of the image slicer made the efficiency of UVES 
critically dependent on the seeing, but it is noted that on one night with
excellent seeing on Paranal (0.35 arcsec) the efficiency of UVES was as high as
could be expected in the image slicer mode for average seeing conditions.

The red arm of UVES was applied with the
two CCDs (EEV $4096 \times 2048$ and MIT-LL $4096 \times 2048$) covering
the spectral regions 610 - 705 and 715 - 810 nm in 14 and 10 echelle orders,
respectively. Cross disperser \#3 was chosen to
provide ample space ($\simeq 150$ pixels) between the orders
for accurate measurement of the background.

\begin{figure}[t]
\resizebox{\hsize}{!}{\includegraphics{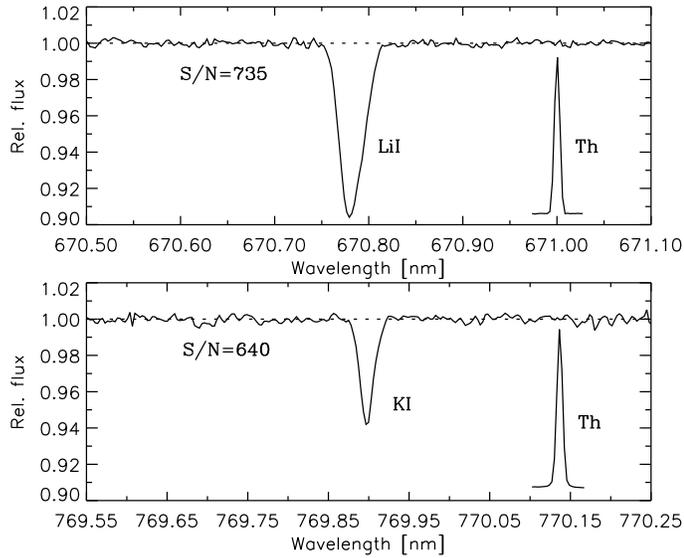}}
\caption{UVES spectrum of \object{G\,271-162}
in the regions of the \Lione\,670.8
and \Kone\,769.9\,nm lines with nearby thorium emission lines inserted}
         \label{f:LiK}
\end{figure}

The spectra have been reduced using standard IRAF tasks for
order definition, background subtraction, FF division, extraction of
orders, wavelength calibration  and continuum normalization. The spectrum of
a bright B-type star, \object{HR\,8858}, was observed and used to remove
telluric lines present in some spectral regions.
After correction for radial velocity shifts
the four spectra were combined to one. 
Fig. \ref{f:LiK} shows the very high $S/N$ and resolution obtained. 

The wavelength calibration is based on  30-40 thorium lines per 
echelle order, and the rms of a 2-dimensional 
4th order polynomial fit between pixel and
wavelength space is 1.1\,m\AA . In addition, the Th 
lines can be used to check the resolution; the instrument profile
is well approximated with a Gaussian and no asymmetry is
seen. The resolution as measured from the FWHM of the Th lines varies, however,
along a given spectral order from about 100\,000
to 120\,000. Thanks to the large number of Th lines available, the variation
could be mapped and the actual instrumental resolution for a given stellar
line determined and applied in connection with the spectrum synthesis.
It is noted that this problem of resolution variations has been
fixed during the second commissioning run of UVES.

\section{Analysis}

The analysis of the \Lione\ feature (equivalent width 29.0\,m\AA )
has been carried out using both
1D hydrostatic (Asplund et al. \cite{asplund97})
and 3D hydrodynamical LTE model atmospheres (Asplund et al. \cite{asplund99}). 
According to Str\"omgren photometry the 
stellar parameters of \object{G\,271-162} are
$\teff = 6295 \pm 70$\,K and $\feh = -2.15\pm 0.2$ while the uncertain
Hipparcos parallax suggests $\logg = 3.7^{+0.4}_{-0.7}$.
\object{G\,271-162} closely resembles \object{HD\,84937}
for which the IR flux method (IRFM),
the Hipparcos parallax and stellar spectroscopy imply
$\teff = 6330 \pm 80$\,K, $\logg = 4.0\pm0.1$ and 
$\feh = -2.25\pm0.2$, which we therefore also adopt for \object{G\,271-162}. It
should be emphasized though that the exact choice of parameters does not
influence significantly the lithium {\em isotope ratio} determinations although
it does of course affect the {\em absolute} abundances.

When applying 1D model atmospheres to spectral synthesis additional
line broadening besides the instrumental broadening is required.
This broadening, which is mainly due to macroturbulence, was approximated
by a Gaussian function and determined from a 
$\chi^2$ fit of five unblended lines 
(\Caone\,612.2, \Caone\,616.2, \Feone\,623.0, \Caone\,643.9, and
\Kone\,769.9\,nm), all of similar strength as the \Lione\,670.8\,nm line.
In the case of the \Kone\,769.9\,nm resonance line the hyperfine structure
and isotopic shift between \Kxxx\ and \Kyyy\ were taken into account
although the effect turned out to be rather negligible.
A more
sophisticated broadening like a radial-tangential profile did not improve
the agreement with the observed profiles. In contrast, in hydrodynamical
model atmospheres the self-consistent
convective motions account for all of the line
broadening except for the unknown rotational velocity \vsini\ of the star.
This parameter was determined by fitting a 
disk-integration of the angle-dependent theoretical 3D line profiles
to the observed profiles of the \Kone, \Caone\ and \Feone\ lines.

\begin{figure}[t]
\resizebox{\hsize}{!}{\includegraphics{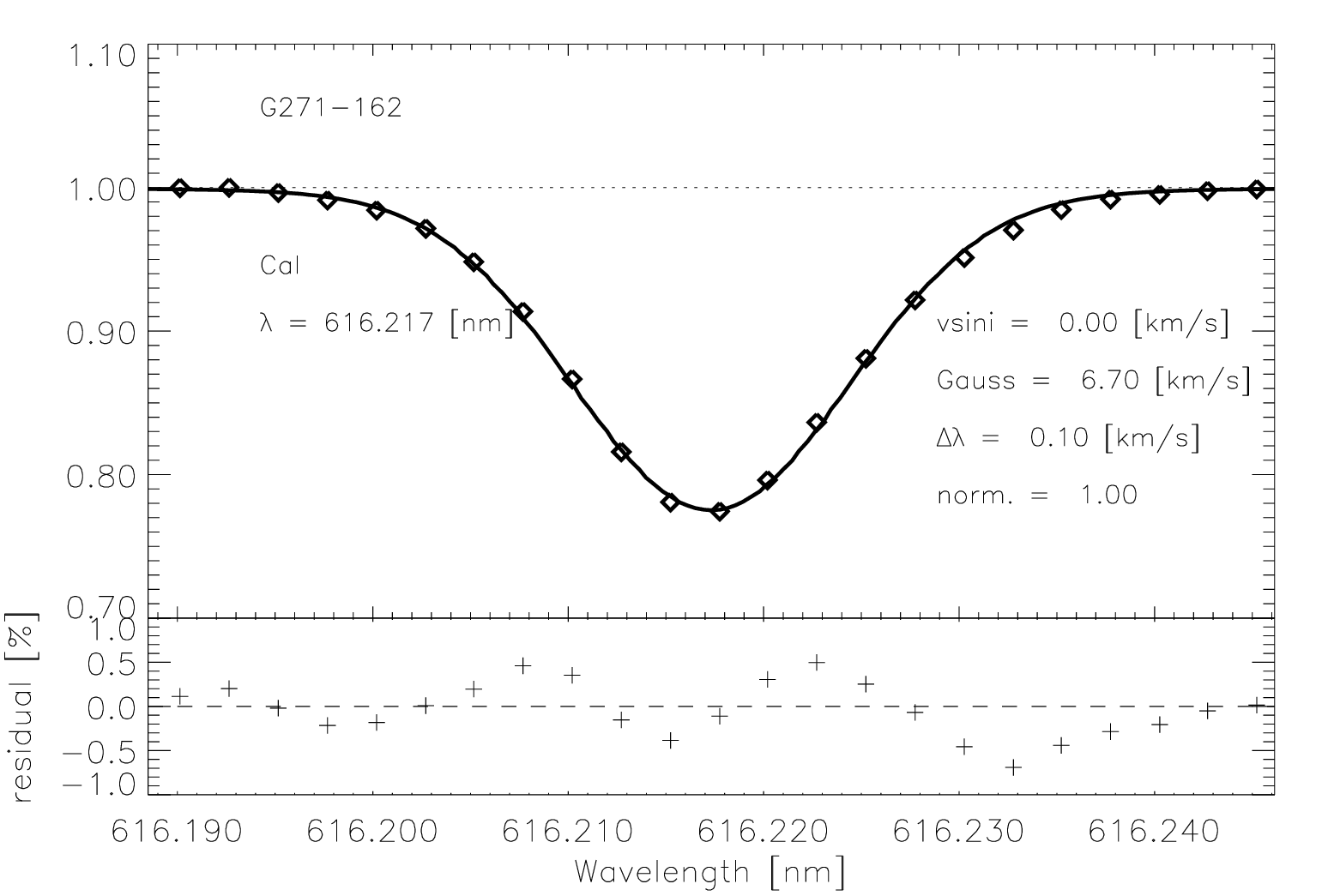}}
\caption{The predicted (solid line) and observed (diamonds) 
\Caone\,616.2\,nm line together with the residual fluxes based 
on the 1D analysis. The Gaussian broadening with which the theoretical
spectrum has been convolved include the combined effects of
macroturbulence, rotation and instrumental resolution.
Note the remaining systematic signal in the residuals}
         \label{f:ca1d}
\end{figure}

\begin{figure}[t]
\resizebox{\hsize}{!}{\includegraphics{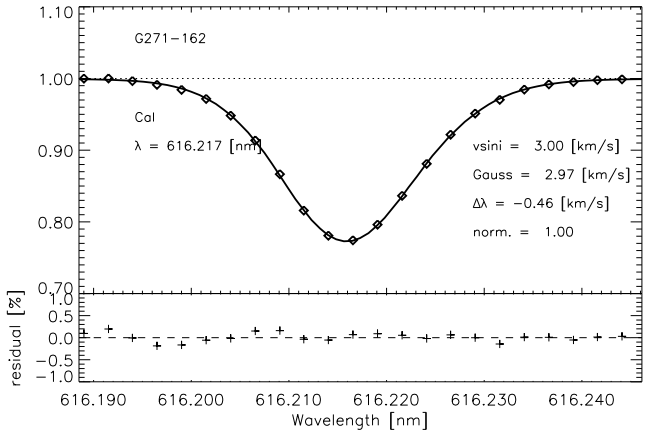}}
\caption{The predicted (solid line) and observed (diamonds) 
\Caone\,616.2\,nm line together with the residual fluxes based 
on the 3D analysis. The rotational broadening has been included through
a disk-integration after which a Gaussian instrumental profile
has been applied. Note the excellent ability of the
3D profile to describe the observed red asymmetry of the line not
possible with 1D models}
         \label{f:ca3d}
\end{figure}

The comparison between theoretical and observed profiles was quantified
using a $\chi^2$ analysis in an analogous way to previous investigations
(Nissen et al. \cite{nissen99}). The $\chi^2$ was computed through
$\chi^2 = \sum (O_i - S_i)^2/\sigma^2$, 
where $O_i$ is the observed spectral point, $S_i$ the synthesis and
$\sigma = (S/N)^{-1}$ as estimated in two adjacent continuum 
windows to the lines.
When estimating the FWHM of the Gaussian broadening (1D) and \vsini\ (3D),
the abundance and the wavelength zeropoint of the observed spectrum
were allowed to vary to
optimize the fit (Figs. \ref{f:ca1d} and \ref{f:ca3d}). 
The most probable value corresponds to the minimum
of $\chi^2$ and $\Delta \chi^2 = \chi^2 - \chi^2_{\rm min} = 1, 4$ and 9
represent the 1-, 2- and 3$\sigma$ confidence limits of the determinations.
These broadening parameters were subsequently used in a similar 
$\chi^2$ analysis
of the \Lione\,670.8\,nm line. The free parameters in the comparison
between the predicted and observed profile were, besides
the \Lisix\ fraction, 
%($\sixseven = N(\Lisix)/N({\rm Li})$),
the total Li abundance and the wavelength zeropoint of the observed profile.
It is emphasized that for each value of \sixseven\ the other
free parameters are optimized to get the smallest possible value
of $\chi^2$.

The 1D analysis of the \Kone, \Caone\ and \Feone\ lines 
implies a $FWHM = 5.9 \pm 0.1$\,\kmprs of the stellar line broadening function.
However, the detailed 
agreement with the observations are far from perfect (Fig. \ref{f:ca1d}). 
In particular,
the distinct red asymmetry of all lines similar to the solar case
(e.g. Asplund et al. \cite{asp00}) is not accounted for.
The result of the $\chi^2$ analysis of the \Lione\,670.8\,nm doublet
suggests a possible detection of \Lisix: $\sixseven = 0.02\pm0.01$ ($1\sigma$).
The minimum $\chi_{\rm red}^2 = \chi^2/\nu$, where $\nu$ is the number 
of degrees of freedom in the fit, is satisfactory close to 1.

\begin{figure}[t]
\resizebox{\hsize}{!}{\includegraphics{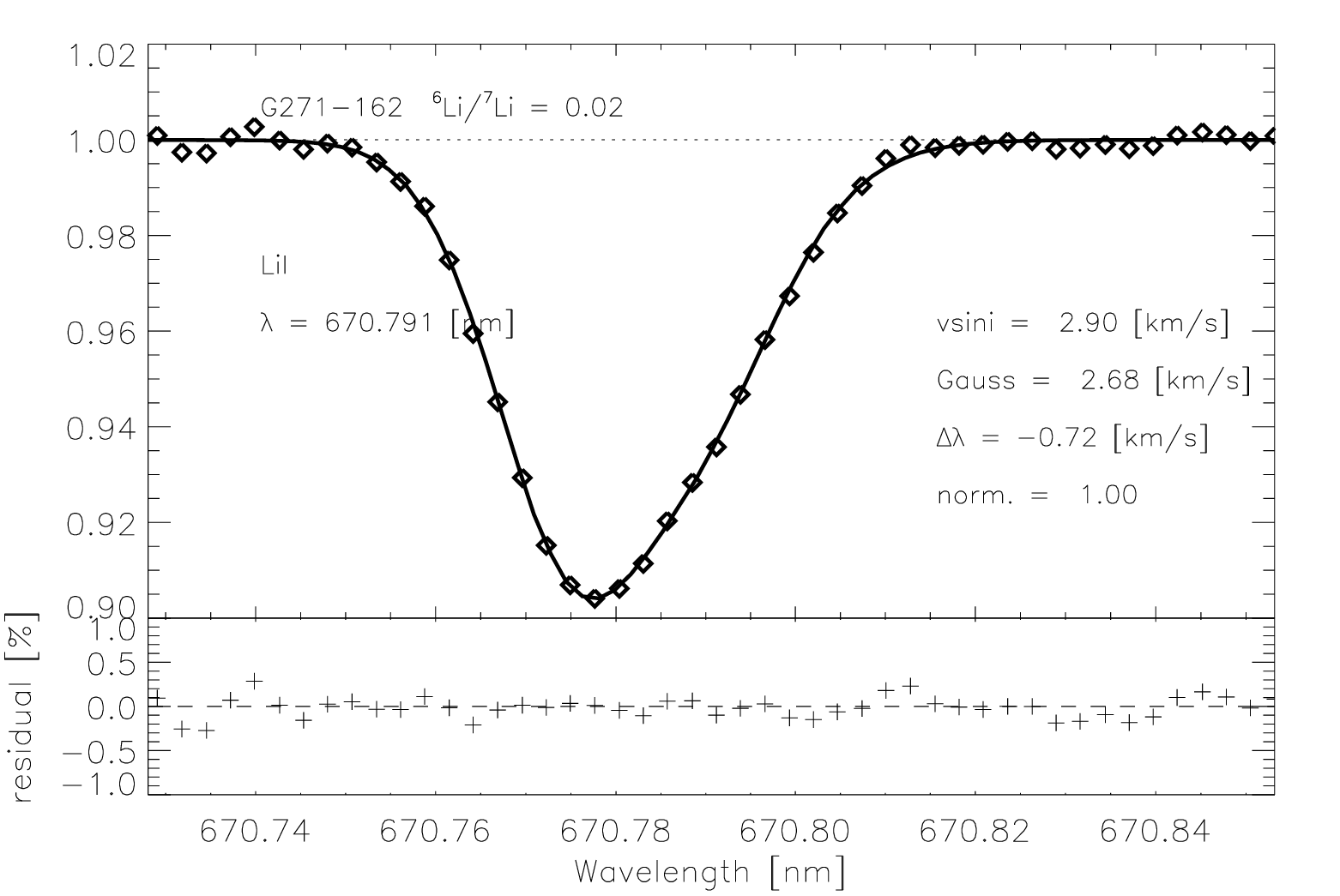}}
\caption{The predicted (solid line) and observed (diamonds) 
\Lione\,670.8\,nm lines together with the residual fluxes for
the best fit with $\sixseven = 0.02$ in the 3D analysis. The Gaussian 
broadening with which the theoretical
spectrum has been convolved include only the instrumental resolution while
the rotational broadening has been applied following a full disk-integration
of the 3D profiles}
         \label{f:li3d}
\end{figure}

\begin{figure}[t]
\resizebox{\hsize}{!}{\includegraphics{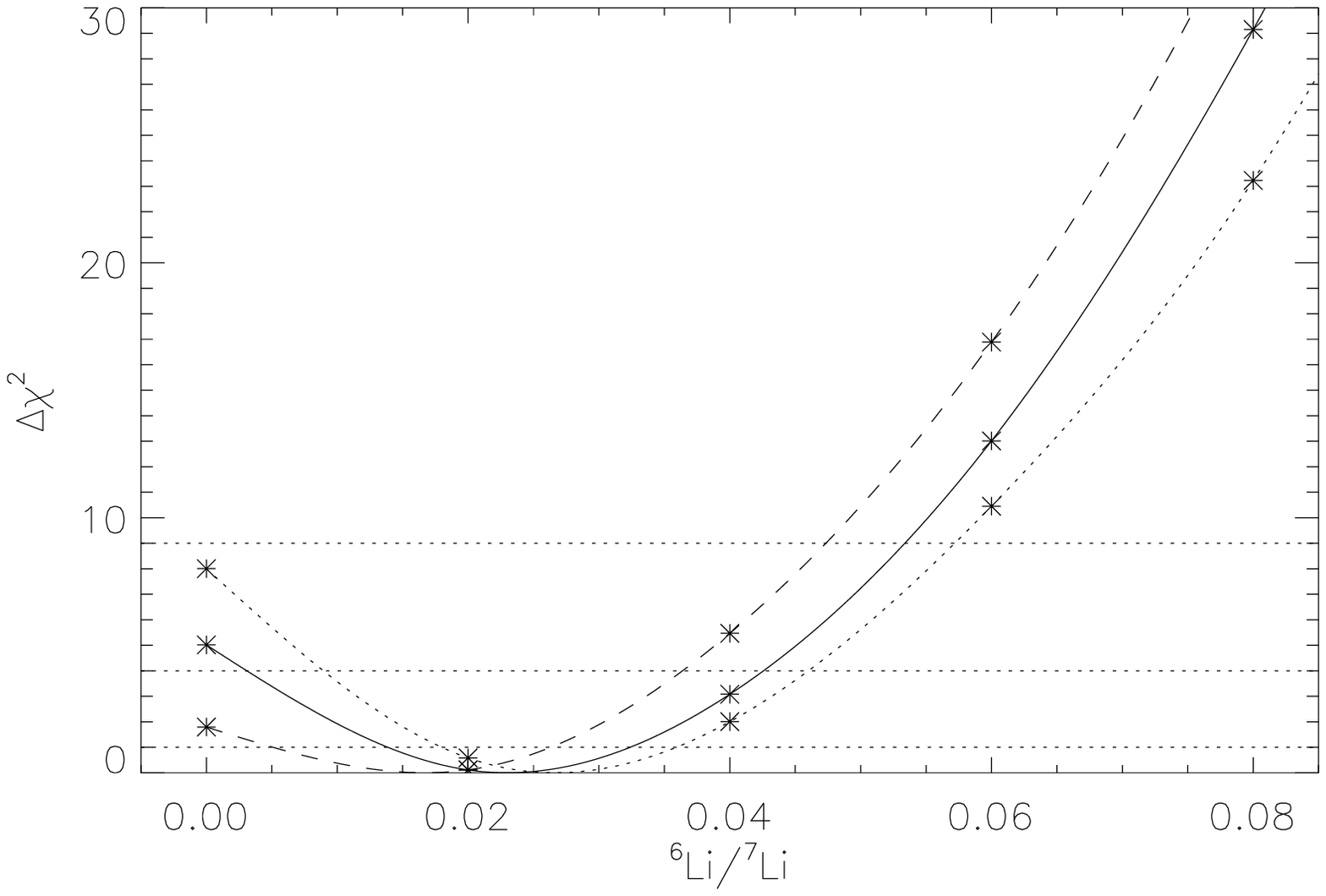}}
\caption{The variation of $\Delta \chi^2$ in the 3D analysis of the 
\Lione\,670.8\,nm line as a function of the relative abundance
of \Lisix\ for different adopted \vsini: 
2.7 (dotted), 2.9 (solid) and 3.1\,\kmprs\ (dashed). 
The corresponding estimates of \sixseven\ are 
$0.027\pm0.009$, $0.023\pm0.009$ and
$0.017\pm0.010$, respectively. The widths of \Kone, \Caone\ and \Feone\ lines
imply $\vsini = 2.9\pm0.1$\,\kmprs. The horizontal lines represent
the formal $1\sigma$, $2\sigma$ and $3\sigma$ confidence limits }
         \label{f:chi3d}
\end{figure}

The temperature inhomogeneities and velocity fields in the
3D hydrodynamical model atmospheres introduce characteristic line
asymmetries, which for the Sun perfectly describe the observed
asymmetries even on an absolute wavelength scale
(Asplund et al. \cite{asp00}). The existence of the predicted
line asymmetries in the 3D analysis vastly improves the agreement
with the observed profiles for \object{G\,271-162} (Fig. \ref{f:ca3d}); other
lines show similarly encouraging correspondence.
This provides additional support for the 
realism of the 3D model when describing the real
stellar photosphere, as well as giving a higher degree of confidence in
the estimate of the lithium isotope ratio. 
The stellar rotation was determined to be $\vsini\ = 2.9\pm0.1$\,\kmprs ,
which in turn implies $\sixseven = 0.02\pm0.01$ ($1\sigma$) according to
Fig \ref{f:chi3d}, the same result as in 
the 1D analysis. Again, the minimum $\chi_{\rm red}^2$ is close to 1
as it should.

\section{Discussion and conclusions}
Both the 1D and 3D analyses suggest a possible 
detection of \Lisix\ in \object{G\,271-162} at the level of 
$\sixseven\ = 0.02\pm0.01$. Given the remaining uncertainties in
the determination of the stellar line broadening (e.g. NLTE effects)
the detection should, however, be considered preliminary.
The similarity between the 1D and 3D results and the excellent
agreement between the observed and predicted profiles in 3D,
could be interpreted as support for a positive detection though.
It also lends added confidence to previous detections of 
\Lisix\ in metal-poor stars based on 1D investigations
(e.g. Smith et al. \cite{smith98}; Nissen et al. \cite{nissen99}).
Evidently, the convective
line asymmetries are rather negligible compared to the isotopic shift
of the \Lisix\ doublet.

Regardless of whether \Lisix\ has been detected in \object{G\,271-162}
or not, the \sixseven\ ratio appears in any case to be smaller
than in \object{HD\,84937} 
for which a weighted average of the results of Hobbs \& Thorburn 
(\cite{hobbs97}), Smith et al. (\cite{smith98}) 
and Cayrel et al. (\cite{cayrel99}) is
$\sixseven = 0.059 \pm 0.016 \,\, (1\sigma)$. This is quite puzzling,
because the two stars have almost identical parameters according to the 
$uvby$-$\beta$ photometry of Schuster \& Nissen (\cite{schuster88}).
Table 1 lists
the measured indices for the three halo stars with \Lisix\ measurements
after correction for a small reddening of G\,271-162,
$E(b-y) = 0.020$, as derived from the the $\beta$ - $(b-y)_0$ calibration
of Schuster \& Nissen (\cite{schuster89}). Table 2 shows the derived parameters
using the IRFM calibration of \teff\ vs. $(b-y)_0$ by
Alonso et al. (\cite{alonso96a}),
the \feh\ calibration of Schuster \& Nissen (\cite{schuster89}) and
the $M_V$ calibration of Nissen \& Schuster (\cite{nissen91}).
The directly measured IRFM temperatures of 
\object{HD\,84937} and \object{BD\,+26\,3578} are 6330 and 6310
(Alonso et al. \cite{alonso96b}) in excellent 
agreement with the values in Table 2, whereas \object{G\,271-162}
has not been measured. 
The absolute magnitude of \object{HD\,84937} from the Hipparcos parallax 
(ESA, \cite{esa97}) is
$M_V = 3.82 \pm 0.19$ in good agreement with the photometric value within
the quoted errors. For the two other stars the relative error of the
Hipparcos parallax is far too large to estimate $M_V$ with any reasonable
accuracy. Finally, a preliminary abundance analysis shows that
\object{HD\,84937} and \object{G\,271-162}
have the same metal abundances within $\pm 0.15$~dex.

\begin{table}
\caption[]{Str\"{o}mgren photometry from Schuster \& Nissen (\cite{schuster88})}
\begin{tabular}{lrcccc}
\hline\noalign{\smallskip}
 ID & $V$ &  $(b-y)_0$ & $m_0$ & $c_0$ & $\beta$  \\
\noalign{\smallskip}
\hline\noalign{\smallskip}
 HD\,84937   & 8.33 & 0.303 & 0.056 & 0.354 & 2.613      \\
 G\,271-162  &10.35 & 0.306 & 0.055 & 0.355 & 2.602      \\
 BD\,+26\,3578& 9.37 & 0.308 & 0.045 & 0.366 & 2.600      \\
\noalign{\smallskip}
\hline
\end{tabular}
\end{table}

\begin{table}
\caption[]{Stellar parameters and Li isotope ratios} 
%The estimated errors
%are $\sigma(\teff) = \pm 70$~K, $\sigma (\feh) = \pm 0.20$, and 
%$\sigma (M_V) = \pm 0.25$
\begin{tabular}{lcccc}
\hline\noalign{\smallskip}
 ID & $\teff$ & \feh &  $M_V$ & \sixseven    \\
\noalign{\smallskip}
\hline\noalign{\smallskip}
 HD\,84937   & 6315\,K  & $-2.14$ &  3.58  & $0.059 \pm 0.016$ \\
 G\,271-162  & 6295\,K  & $-2.15$ &  3.53  & $0.020 \pm 0.010$ \\
 BD\,+26\,3578& 6280\,K  & $-2.60$ &  3.08  & $0.050 \pm 0.030$ \\
\noalign{\smallskip}
\hline
\end{tabular}
\end{table}

\begin{figure}[t]
\resizebox{\hsize}{!}{\includegraphics{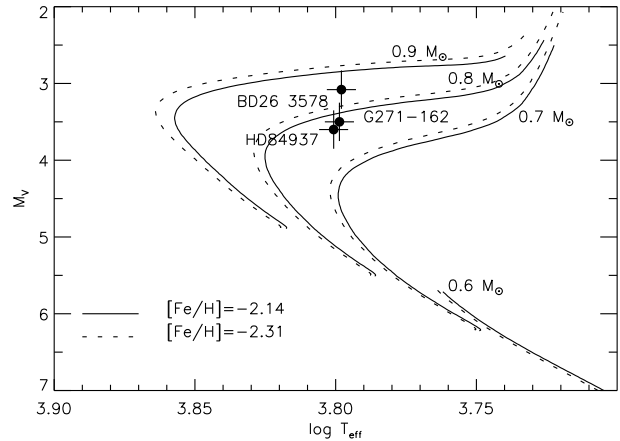}}
\caption{The position of \object{HD\,84937}, \object{BD+26\,3578}
and \object{G\,271-162} in
the \mvlgt\ diagram compared to evolutionary tracks from VandenBerg et al. 
(\cite{vandenberg00}) with masses and metallicities indicated}
         \label{f:HR}
\end{figure}

Fig. \ref{f:HR} shows the location of the three stars in the
\mvlgt\ diagram.
Clearly, \object{HD\,84937} and \object{G\,271-162} have nearly the same mass.
This makes it difficult to explain the lower \Lisix\ abundance in 
\object{G\,271-162} as a depletion effect and raises the interesting question
of a possible intrinsic scatter of \sixseven\ in the ISM
at a given metallicity.  In this connection we note that some models
for the formation of the light elements by cosmic ray processes in the
early Galaxy predict a scatter of one order of magnitude in the
abundance of \Lisix , Be and B relative to Fe, e.g. the bimodal superbubble
model of Parizot \& Drury (\cite{parizot99}) and the supernovae-driven chemical
evolution model for the Galactic halo by Suzuki et al. (\cite{suzuki99}). 

It is clear that the ability of UVES on VLT/Kueyen 
to gather high quality, high resolution and very
high $S/N$ spectra of even $V \simeq 10\fm5$ stars 
as evident from the present
study, has opened up the possibility for a
large survey of \sixseven\ for stars with a wide range of metallicities.
An observing program with 
this exact aim using UVES/VLT is currently in progress, which
is expected to be of great importance both for an improved understanding
of Big Bang nucleosynthesis, cosmic chemical evolution of the light
elements and stellar mixing events.

\begin{acknowledgements}
We are greatly indebted to all the people involved in the
conception, construction and commissioning of the UVES instrument, 
without whom this project would have been impossible.
\end{acknowledgements}

\end{document}